\documentstyle[aps,epsfig,twocolumn]{revtex}

\begin{document}
\title{Efficient quantum computation with probabilistic quantum gates}
\author{L.-M. Duan$^{1}$ and R. Raussendorf $^{2}$}
\address{$^{1}$ FOCUS Center and MCTP, Department of Physics,
University of Michigan, Ann Arbor, MI 48109-1120\\
$^{2}$Institute for quantum information, California Institute of
Technology, Pasadena, CA 91125} \maketitle

\begin{abstract}
With a combination of the quantum repeater and the cluster state approaches,
we show that efficient quantum computation can be constructed even if all
the entangling quantum gates only succeed with an arbitrarily small
probability $p${\em {.}} The required computational overhead scales
efficiently both with $1/p$ and $n$, where $n$ is the number of qubits in
the computation. This approach provides an efficient way to combat noise in
a class of quantum computation implementation schemes, where the dominant
noise leads to probabilistic signaled errors with an error probability $1-p$
far beyond any threshold requirement.
\end{abstract}

The celebrated threshold theorem in quantum computation has assured that if
the amount of noise per quantum gate is less than a small value, reliable
quantum computation can be efficiently performed \cite{1}. In terms of
implementation, however, the experimental noise is typically orders of
magnitude large than the required threshold value. To overcome this problem,
a practical route to noise reduction is by exploitation of certain
properties of the noise. In carefully designed implementation schemes, the
dominant noise only leads to specific types of errors which can be corrected
much more efficiently. Here, we consider an important noise model, which is
relevant for several experimental approaches to quantum computation \cite
{2,2',3,4,5,6,7,8,9,10}. In this model, the dominant noise only leads to
significant failure probability for the entangling gates, and a gate failure
is always signaled through a built-in photon or atom detection during the
gate operation. The success probability $p$ for each entangling gate is
rather small for some typical experimental systems \cite{3,9,10}. It is hard
to use the standard methods of error correction in the considered scenario,
because the error probability $1-p$ (close to the unity) is simply too large.

Naively, if a gate only succeeds with a certain probability $p$, one cannot
have efficient computation as the overall success probability (efficiency)
scales down exponentially as $p^{n}$ with the number $n$ of gates. However,
in this paper we show that efficient quantum computation can be constructed
with the required computational overhead (such as the computation time or
the repetition number of the entangling gates) scaling up slowly
(polynomially) with both $n$ and $1/p$. The demonstration of this result
combines the ideas from the quantum repeater schemes \cite{11,12} and the
cluster state approach to computation \cite{13,14}. It has been shown in
quantum repeater schemes that with probabilistic entangling operations, one
can construct scalable quantum communication and GHZ correlations \cite
{12,15}. Recently, it has also been demonstrated in the context of linear
optics computation \cite{2}\ that the threshold requirement on the
probability $p$ for the entangling gates can be significantly improved using
the cluster state approach to quantum computation \cite{2',4,6,7}. In
particular, Ref. \cite{7} shows that for construction of one-dimensional
(1D) cluster states, to get efficient scaling with $n$, in principle no
threshold value is needed on $p$, although in practice $p$ is still required
to be sufficiently large as the computational overhead in that scheme has
very inefficient (super-exponential) scaling with $1/p$. Compared with these
previous results, here we have the following two advances: (i) we propose a
probabilistic computation scheme which has efficient scaling with both $n$
and $1/p$. This improvement is substantial as in current experiments $1/p$
is large \cite{9,10}. (ii) Through explicit construction, we also
demonstrate efficient scaling of the computational overhead for generation
of the two-dimensional (2D) cluster states which are critical for
realization of universal quantum computation.

To be more specific, we assume in this paper that one can reliably perform
two-qubit controlled phase flip (CPF) gates with a small success probability
$p$, although the basic ideas here also apply for other kinds of entangling
gates. We neglect the noise for all the single-bit operations, which is well
justified for typical atomic or optical experiments. Our basic steps are:
first we show how to efficiently prepare 1D cluster state from probabilistic
CPF gates, then we give a construction to efficiently generate 2D cluster
states from 1D chains. Efficient preparation of 2D cluster states, together
with simple single-bit operations, realizes universal quantum computation.

With respect to a given lattice geometry, the cluster state is defined as
co-eigenstates of all the operators $A_{i}=X_{i}\prod_{j}Z_{j}$, where $i$
denotes an arbitrary lattice site and $j$ runs over all the nearest
neighbors of the site $i$. The $X_{i}$ and $Z_{j}$ denote respectively the
Pauli spin and phase flip operators on the qubits at the sites $i,j$. In our
construction of lattice cluster states with probabilistic CPF gates, we will
make use of the following three properties of the cluster states: (i) If we
have two chains of cluster states each with $n$ qubits, we can join them to
form a 1D cluster state of $2n$ qubits by successfully applying a CPF gate
on the end qubits of the two chains. (ii) If we destroy the state of an end
qubit of an $n$-qubit cluster chain, for instance, through an unsuccessful
attempt of the CPF gate, we can remove this bad qubit by performing a $Z$
measurement on its neighboring qubit, and recover a cluster state of $n-2$
qubits. (iii) We can shrink a cluster state by performing $X$ measurements
on all the connecting qubits (see Fig. 1c). These three properties of the
cluster states, illustrated in Fig. 1, can be conveniently explained from
their above definition \cite{14,16}.

\begin{figure}[tbp]
\epsfig{file=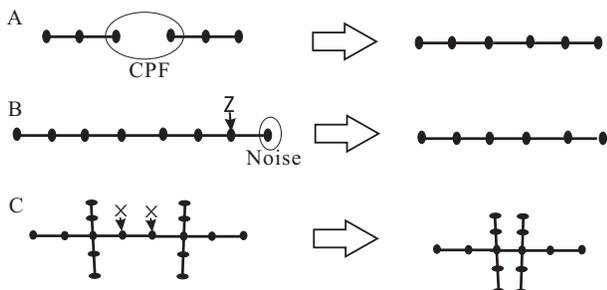,width=8cm}
\caption[Fig.1 ]{Illustration of the three properties of the
cluster states which are important for our construction of such
states with the probabilistic entangling gates: (A) extend cluster
states with CPF gates; (B) recover cluster states by removing bad
qubits; (C) shrink cluster states for more complicated links.}
\end{figure}

If we have generated two sufficiently long cluster chains each of $n_{0}$
qubits, we can just try to connect them through a probabilistic CPF gate. If
this attempt fails, through the property (ii), we can recover two $\left(
n_{0}-2\right) $-qubit cluster chains through a $Z$-measurement, and try to
connect them again. As one continues with this process, the average number
of the qubits in the connected chain is then given by $n_{1}=%
\sum_{i=0}^{n_{0}/2}2\left( n_{0}-2i\right) p\left( 1-p\right) ^{i}\simeq
2n_{0}-4\left( 1-p\right) /p$, where the last approximation is valid when $%
e^{-n_{0}p/2}\ll 1$. So the average chain length goes up if $%
n_{0}>n_{c}\equiv 4\left( 1-p\right) /p$. We can iterate these connections
to see how the computation overhead scales with the qubit number $n$. We
measure the computation overhead in terms of the total computation time and
the total number of attempts for the CPF\ gates. For the $r$th $\left( r\geq
1\right) $ round of successful connection, the chain length $n_{r}$, the
total preparation time $T_{r}$, and the total number of attempts $M_{r}$
scale up respectively by the recursion relations $n_{r}=2n_{r-1}-n_{c}$, $%
T_{r}=T_{r-1}+t_{a}/p$, $M_{r}=2M_{r-1}+1/p$. In writing the recursion
relation for $T_{r}$, we have assumed that two cluster chains for each
connection are prepared in parallel, and we neglect the time for single-bit
operations ($t_{a}$ denotes the time for each attempt of the CPT\ gate).
From the above recursion relations, we conclude that if we can prepare
cluster chains of $n_{0}$ $\left( n_{0}>n_{c}\right) $ qubits in time $T_{0}$
with $M_{0}$ attempts of the probabilistic gates, for a large cluster state,
the preparation time $T$ and the number of attempts $M$ scale with the chain
length $n$ as $T\left( n\right) =T_{0}+\left( t_{a}/p\right) \log _{2}\left[
\left( n-n_{c}\right) /\left( n_{0}-n_{c}\right) \right] ,$ and $M\left(
n\right) =\left( M_{0}+1/p\right) \left( n-n_{c}\right) /\left(
n_{0}-n_{c}\right) -1/p$.

In the above, we have shown that if one can prepare cluster chains longer
than some critical length $n_{c}$, one can generate large scale 1D cluster
states very efficiently. The problem then reduces to how to efficiently
prepare cluster chains up to the critical length $n_{c}$. If one wants to
prepare an $n$-qubit cluster chain, we propose to use a repeater protocol
which divides the task into $m=\log _{2}n$ steps: for the $i$th $(i=1,2,..,m)
$ step we attempt to build a $2^{i}$-bit cluster state by connecting two $%
2^{i-1}$-bit cluster chains through a probabilistic CPF gate. If such an
attempt fails, we discard all the qubits and restart from the beginning \cite
{note}. For the $i$th step, the recursion relations for the preparation time
$T_{i}$ and the number of attempts $M_{i}$ are given by $T_{i}=(1/p)\left(
T_{i-1}+t_{a}\right) $ \cite{note2} and $M_{i}=(1/p)\left( 2M_{i-1}+1\right)
$, which, together with $T_{1}=t_{a}/p$ and $M_{1}=1/p$, give the scaling
rules $T\left( n\right) \simeq t_{a}\left( 1/p\right) ^{\log _{2}n}$ and $%
M\left( n\right) \simeq \left( 2/p\right) ^{\log _{2}n}/2$. The cost is more
significant, but it is still a polynomial function of $n$. To construct a $n$%
-qubit cluster chain, in total we need $n-1$ successful CPF gates. In a
direct protocol, we need all these attempts succeed simultaneously, which
gives the scaling $T\left( n\right) \propto M\left( n\right) \propto
(1/p)^{n-1}$. By dividing the task into a series of independent pieces, we
improve the scaling with $n$ from exponential to polynomial (for $n\leq n_{c}
$).

To generate a cluster chain of a length $n>n_{c}$, we simply combine the
above two protocols. First, we use the repeater protocol to generate $n_{0}$%
-qubit chains with $n_{0}>n_{c}$. Then it is straightforward to use the
connect-and-repair protocol to further increase its length. For instance,
with $n_{0}=n_{c}+1$ (which is a reasonable close-to-optimal choice), the
overall scaling rules for $T$ and $M$ are (for $n>n_{c}$),
\begin{equation}
T\left( n\right) \simeq t_{a}\left( 1/p\right) ^{\log _{2}\left(
n_{c}+1\right) }+\left( t_{a}/p\right) \log _{2}\left( n-n_{c}\right) ,
\end{equation}
\begin{equation}
M\left( n\right) \simeq \left( 2/p\right) ^{\log _{2}\left( n_{c}+1\right)
}\left( n-n_{c}\right) /2.
\end{equation}
As the critical length is $n_{c}\simeq 4/p$, $T$ and $M$ in our protocol
scale with $1/p$ as $\left( 1/p\right) ^{\log _{2}\left( 4/p\right) }$,
which is much more efficient than the super-exponential scaling $\left(
1/p\right) ^{4/p}$ in the previous work \cite{7}.

We have shown that for any success probability $p$ of the probabilistic
entangling gate, 1D cluster states of arbitrary length can be created
efficiently. For universal quantum computation, however, such 1D cluster
states are not sufficient. They need to be first connected and transformed
into 2D cluster states (for instance, with a square lattice geometry) \cite
{14}. It is not obvious that such a connection can be done {\it efficiently}%
. First, in the connect-and repair protocol, when an attempt fails, we need
to remove the end qubits and all of their neighbors. This means that in a 2D
geometry, the lattice shrinks much faster to an irregular shape in the
events of failure. Furthermore, a more important obstacle is that we need to
connect much more boundary qubits if we want to join two 2D cluster states.
For instance, for a square lattice of $n$ qubits, the number of boundary
qubits scales as $\sqrt{n}$ (which is distinct from a 1D chain). If we need
to connect all the corresponding boundary qubits of the two parts, the
overall success probability is exponentially small.

To overcome this problem, we introduce a method which enables efficient
connection by attaching a long leg (a 1D cluster chain) to each boundary
qubit of the 2D lattice. The protocol is divided into the following steps:
First, we try to build a ``+'' shape cluster state by probabilistically
connecting two cluster chains each of length $2n_{l}+1$ (the value of $n_{l}$
will be specified below). This can be done through the probabilistic CPF
gate together with a simple Hardmard gate $H$ and an $X$-measurement, as
shown in Fig. 2A and explained in its caption. With on average $1/p$
repetitions, we get a ``+'' shape state with the length of each of the four
legs given by $n_{l}$. We use the ``+'' shape state as the basic building
blocks of large scale 2D cluster states. In the ``+'' shape state, we have
attached four long legs to the center qubit. The leg qubits serve as ancilla
to generate near-deterministic connection from the probabilistic CPF\ gates.
The critical idea here is that if we want to connect two center qubits, we
always start the connection along the end qubits of one of the legs (see
illustration in Fig. 2). If such an attempt fails, we can delete two end
qubits and try the connection again along the same legs. If the leg is
sufficiently long, we can almost certainly succeed before we reach (destroy)
the center qubits. When we succeed, and if there are still redundant leg
qubits between the two center ones, we can delete the intermediate leg
qubits by performing simple single-bit $X$ measurements on all of them (see
Fig. 2 and Fig. 1C for the third property of the cluster state). With such a
procedure, we can continuously connect the center qubits and form any
complex lattice geometry (see the illustration for construction of the
square lattice state in Fig. 2B and 2C). What is important here is that
after each time of connection of the center qubits, in the formed new shape,
we still have the same length of ancillary legs on all the boundary qubits,
which enables the succeeding near-deterministic connection of these new
shapes.

\begin{figure}[tbp]
\epsfig{file=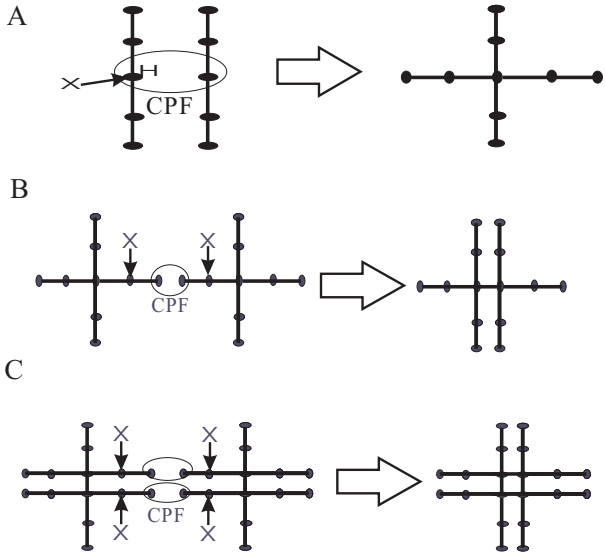,width=8cm}
\caption[Fig.2 ]{Illustration of the steps for construction of the
two-dimensional square lattice cluster states from a set of
cluster chains. (A) Construction of the basic ``+" shape states
from cluster chains by applying first a Hardmard gate H on the
middle qubit of one chain, and then a CPF gate to connect the two
middle qubits, and finally a $X$ measurement on one middle qubit
to remove it. (B,C) Construction of the square lattice cluster
state from the ``+" shape states through probabilistic CPF gates
along the legs and $X$ measurements to remove the remaining
redundant qubits. }
\end{figure}

Now we investigate for the 2D case how the computational overhead scales
with the size of the cluster state. If the ancillary legs have length $n_{l}$%
, for each connection of two center qubits, we can try at most $n_{l}/2$
times of the probabilistic CPF gates, and the overall success probability is
given by $p_{c}=1-\left( 1-p\right) ^{n_{l}/2}$. If we want to build a
square lattice cluster state of $N$ qubits, we need about $2N$ times of
connections of the center qubits (there are about $2N$ edges in an $N$%
-vertex square lattice). The probability for all these connections to be
successful is given by $p_{c}^{2N}$. We require this overall success
probability is sufficiently large with $p_{c}^{2N}\geq 1-\epsilon $, where $%
\epsilon $ is a small number characterizing the overall failure probability.
From that requirement, we figure out that $n_{l}\simeq \left( 2/p\right) \ln
\left( 2N/\epsilon \right) $. To construct a square lattice cluster state of
$N$ qubits, we need to consume $N$ ``+'' shape states, and each of the
latter requires on average $2/p$ cluster chains with a length of $2n_{l}+1$
qubits. So we need in total $2N/p$ $\left( 2n_{l}+1\right) $-bit\ cluster
chains, which can be prepared in parallel with $\left( 2N/p\right)
M(2n_{l}+1)$ CPF\ attempts within a time period $T(2n_{l}+1)$ (see Eqs. (1)
and (2) for expressions of the $M(n)$ and $T(n)$). This gives the resources
for preparation of all the basic building blocks (the chains). Then we need
to connect these blocks to form the square lattice. We assume that the
connection of all the building blocks are done in parallel. The whole
connection takes on average $2N/p$ CPF\ attempts, and consumes a time at
most $t_{a}/p\,\ln (2N/\epsilon ).$ Summarizing these results, the temporal
and the operational resources for preparation of an $N$-bit square lattice
cluster state are approximately given by
\begin{equation}
\begin{array}{rcl}
T\left( N\right)  & \simeq  & t_{a}\left( 1/p\right) ^{\log _{2}\left(
4/p-3\right) } \\
& + & \frac{t_{a}}{p}\log _{2}\left( \frac{4}{p}\left[ \ln \left(
2N/\epsilon \right) -1\right] \right)  \\
& + & \frac{t_{a}}{p}\ln \left( 2N/\epsilon \right) ,
\end{array}
\end{equation}
\begin{equation}
M\left( N\right) \simeq \left( 2/p\right) ^{2+\log _{2}\left( 4/p-3\right) }N
\left[ \ln \left( 2N/\epsilon \right) -1\right] +2N/p.
\end{equation}
In the 2D case, the temporal and the operational overheads still have very
efficient scaling with the qubit number $N$, logarithmically for $T\left(
N\right) $ and $N\ln \left( N\right) $ for $M\left( N\right) $. Their
scalings with $1/p$ are almost the same as in the 1D case except an
additional factor of $1/p^{2}$ for $M\left( N\right) $. Through some
straightforward variations of the above method, it is also possible to
efficiently prepare any complicated graph state using probabilistic CPF
gates \cite{16}. This shows that in principle we do not need to impose any
threshold on the success probability of the CPF\ gates for efficient quantum
computation.

Before ending the paper, we would like to add a few remarks on other sources
of noise that have not been taken into account in the above discussions. If
each CPF\ gate has some small additional infidelity error, one might wonder
whether such an error scales up with the large number of attempts $M\left(
N\right) $. That is actually not the case. Most of the CPF attempts have
failed, and all the failed CPF gates have no contribution to the final state
infidelity. In practice, we may be more concerned about the temporal
overhead $T\left( N\right) $ than the operational overhead $M\left( N\right)
$. Each qubit has a finite coherence time and we need to finish all the CPF
attempts within such a time scale. For typical probabilistic entangling
experiments with atoms \cite{9,10}, the time $t_{a}$ for each CPF attempt is
about $100$ns, while the qubit coherence time is usually longer than a
second. If we take the success probability $p\sim 0.1$, Eq.(3) gives $%
T\left( N\right) \sim 1.6\times 10^{5}t_{a}\sim 16$ms for any large $N$ \cite
{note3}, which is still well within the qubit coherence time.

In summary, we have shown that cluster states in any realistic dimension can
be generated using probabilistic CPF\ gates with efficient scaling in both
the qubit number and the inverse of the success probability. This result
opens up a promising prospect to realize efficient quantum computation with
probabilistic entangling gate operations. Such a prospect is relevant for
several experimental systems involving atoms, ions and photons \cite
{2,3,9,10}, with on-going efforts towards probabilistic quantum information
processing.

LMD thanks Chris Monroe for helpful discussions. This work was
supported by the NSF awards (0431476 and EIA-0086038), the ARDA
under ARO contract, the A. P. Sloan Fellowship, and the visitor
program of MCTP.


\begin{references}
\bibitem{note}  If we do not immediately discard all the qubits, but instead
do some repairing and re-connection as described in the previous paragraph,
the protocol will be somewhat more efficient. But for $n\leq n_{r},$ such
repairing has no significant influence on the basic scaling rules.
\bibitem{note2}  In writing the recursion relation for $T_{i}$, we have
assumed that if we fail, we start over from the first step. Here, we
actually assumed that we have only one copy of the system with $n$ qubits
(no ancillary qubits). If we have many copies of such $n$-qubit systems as
ancilla, in principle we can get much more efficient scaling for $T_{i}$
with parallel operations on all the copies. We try the chain connection in
parallel in all these $n$-qubit systems, discard the systems with failure
events, and continue the next-step connection for the systems where all the
previous connections are successful. In this way, one connection (one length
doubling of the qubit chain) takes only one time step $t_{a}$, so we have $%
T_{i}=T_{i-1}+t_{a}$. This recursion relation will lead to a significant
improvement in the total required preparation time: the scaling of $T\left(
N\right) $ (or $T\left( n\right) $) with $1/p$ in Eq. (3) (or (1)) changes
from $\left( 1/p\right) ^{\log _{2}\left( 4/p\right) }$ to $\left(
1/p\right) \log _{2}\left( 4/p\right) $. Of course, the cost for this
improvement is probably too expensive for practical experiments: to prepare $%
n$-qubit cluster states with the above parallel operations, we need to have
roughly $n\left( 1/p\right) ^{\log _{2}\left( 4/p\right) }$ ancillary
qubits.
\bibitem{note3}  For the quoted number, $T\left( N\right) $ is basically
independent of $N$ for $N/\epsilon <10^{7000}$ due to the slow
logarithmic scaling $\ln \left( 2N/\epsilon \right)$.
\end{references}
\end{document}